\documentclass[notoc,12pt]{JHEP}

\usepackage{amsmath,amssymb,euscript,array,cite}

\input{epsf}
\newcommand{\startappendix}{
\setcounter{section}{0}
\renewcommand{\thesection}{\Alph{section}}}
\newcommand{\Appendix}[1]{
\refstepcounter{section}
\begin{flushleft}
{\large\bf Appendix \thesection: #1}
\end{flushleft}}

\usepackage{epsfig}

\def\N{{\cal N}}

\def\ttau{{\tilde\tau}}

\def\Tr{{\rm Tr}\,}

\def\det{{\rm det}}
\def\SU{\text{SU}}
\def\SO{\text{SO}}
\def\U{\text{U}}

\def\SL{\text{SL}}

\def\Dbarslash{\,\,{\raise.15ex\hbox{/}\mkern-12mu {\bar\D}}}
\def\Dslash{\,\,{\raise.15ex\hbox{/}\mkern-12mu \D}}
\def\delslash{\,\,{\raise.15ex\hbox{/}\mkern-9mu \partial}}
\def\delbarslash{\,\,{\raise.15ex\hbox{/}\mkern-9mu {\bar\partial}}}

\def\Vev#1{\big\langle{#1}\big\rangle}
\def\vev#1{\langle{#1}\rangle}

\newcommand{\MAT}[1]{\begin{pmatrix} #1\end{pmatrix}}
\newcommand{\EQ}[1]{\begin{equation} #1 \end{equation}}
\newcommand{\AL}[1]{\begin{subequations}\begin{align} #1
\end{align}\end{subequations}}
\newcommand{\SP}[1]{\begin{equation}\begin{split} #1 \end{split}\end{equation}}

\title{S-duality of the Leigh-Strassler Deformation via Matrix Models} 
\author{Nick Dorey, Timothy J.~Hollowood and S.~Prem~Kumar\\
Department of Physics, University of Wales Swansea,
Swansea, SA2 8PP, UK\\
E-mail: {\tt t.hollowood@swan.ac.uk}, {\tt n.dorey@swan.ac.uk}
{\tt s.p.kumar@swan.ac.uk}}
\preprint{SWAT-356}
\abstract{We investigate an exactly marginal 
$\N=1$ supersymmetric deformation of $\SU(N)$ 
$\N=4$ supersymmetric Yang-Mills theory discovered by Leigh
and Strassler. We use a matrix model to 
compute the exact superpotential for a further massive
deformation of the $\U(N)$ Leigh-Strassler 
theory. We then show how the exact
superpotential and eigenvalue spectrum for the $\SU(N)$ theory follows by a
process of integrating-in. We find that different vacua are
related by an action of the $\SL(2,{\mathbb Z})$ modular group on the
bare couplings of the theory extending the action of electric-magnetic
duality away from the $\N=4$ theory. We perform non-trivial
tests of the matrix model results against semiclassical field
theory analysis. 
We also show that there are interesting points in
parameter space where condensates can diverge and vacua
disappear. Based on the matrix model results, we propose an exact
elliptic superpotential to describe the theory compactified on a
circle of finite radius.} 

\begin{document}

\section{Introduction}

It has been known for some time that the $\N=4$ supersymmetric
Yang-Mills theory in four dimensions possesses exactly marginal
deformations. Leigh and 
Strassler \cite{LS} identified two $\N=1$ SUSY-preserving exactly marginal
directions other than the $\N=4$ gauge coupling itself,
giving rise to a three (complex) dimensional renormalization-group-fixed 
manifold of $\N=1$ superconformal theories (SCFTs) of which the
$\N=4$ fixed line is a subset.  Given that 
strong and weak coupling on the $\N=4$ fixed line are related
by the action of $\SL(2,{\mathbb Z})$ on $\tau$, the gauge coupling of
the $\N=4$ theory, it is then natural to ask whether this duality extends in
some
non-trivial way over the entire $\N=1$ fixed manifold. 

In this
article we answer this question for a 
particular Leigh-Strassler deformation of the $\N=4$ theory: the
so-called ``$q$-deformation'' \cite{LB}
\footnote{One such deformation has been analysed recently using matrix
models in \cite{berenstein}.}
. By studying a mass
deformation of this Leigh-Strassler SCFT using a recently proposed
connection between $\N=1$ theories and matrix models \cite{DV,mm1,mm2}, we
demonstrate that Montonen-Olive electric-magnetic duality, extended to
the full $\SL(2,{\mathbb Z})$ modular group, extends non-trivially to
this space of $\N=1$ superconformal theories 
accompanied by a well-defined action on the marginal
parameters. Not only will we uncover the duality group action in these
theories, but we will also provide rather powerful checks on the
applicability of the matrix model approach towards solving these
$\N=1$ SUSY theories. (Recent checks of the matrix model proposal for
other $\N=1$ models have been performed in
\cite{berenstein, Fuji:2002wd, ferrari}).  In particular we will test
the superpotential 
and eigenvalue 
spectrum obtained from the matrix model against classical field theory
results and find nontrivial agreement between the two.
Furthermore, rather remarkably the matrix model
results for the vacuum expectation values of the 
effective superpotential can be used to
directly infer an exact, dynamical quantum superpotential for the mass
deformation of the Leigh-Strassler SCFT compactified on ${\mathbb
R}^3\times S^1$. This superpotential written in terms of the 
effective fields of the 3-dimensional theory encodes the entire vacuum
structure of the four-dimensional theory and satisfies extremely
non-trivial checks to be discussed below. It turns out to be a natural
deformation of the elliptic superpotential of \cite{nick} for the mass-deformed
$\N=4$, or $\N=1^*$, theory.  

Relevant deformations of SCFTs can tell us a great deal about
the CFTs themselves; in particular, as is
well-known in the context of $\N=4$ SUSY Yang-Mills and the so-called $\N=2$
elliptic quiver models
\cite{mm2,nick,VW,nickprem,us}, the duality properties of the
parent theories are inherited by the perturbed theories and manifest themselves
as modular
properties of various exactly calculable observables in the
perturbed theories. In fact,  what happens in these known examples is that 
the holomorphic observables in one vacuum get mapped
into the corresponding observables in a different vacuum (realized in
a different phase) 
under the action of $\SL(2,\mathbb Z)$ with well-defined modular
weights. The holomorphic quantities in question can often be computed
exactly and reflect the duality symmetries of the parent theory in
this simple fashion.

The lesson to be drawn from this limited set of examples is
that the consequences of dualities in these SCFTs are readily visible and
more importantly, are readily calculable for certain relevant deformations of
these SCFTs. With this in mind, rather than focusing on the SCFT itself we 
investigate a special mass perturbation of 
the Leigh-Strassler fixed points. This theory is the
$\SU(N)$ gauge theory with $\N=1$ SUSY and three adjoint-valued chiral
multiplets $\hat\Phi^+$, $\hat\Phi^-$ and $\hat\Phi$; the same 
matter content as the $\N=4$ theory, but with the following classical
superpotential\footnote{We choose a normalization in which the
kinetic terms of the chiral multiplets do not have factors of
$1/g_{YM}^2$ in front. In the following, hatted quantities refer to
$\SU(N)$ to distinguish them from $\U(N)$.}
\EQ{W_{\rm cl}=
\Tr\big(i\lambda\hat\Phi[\hat\Phi^+,\hat\Phi^-]_\beta+M\hat\Phi^+\hat\Phi^-
+\mu\hat\Phi^2 \big)\ ,\label{tree}}
where we have defined the $q$-commutator
\EQ{[\hat\Phi^+,\hat\Phi^-]_\beta\equiv\hat\Phi^+\hat\Phi^-e^{i\beta/2}-
\hat\Phi^-\hat\Phi^+e^{-i\beta/2}
}
and where $\lambda$ and $\beta$ are complex bare couplings. 
We also introduce the complex bare 
gauge coupling of this theory $\tau\equiv 4\pi i/g^2_{YM}+\theta/2\pi$.
This superpotential represents a deformation away from the $\N=4$
point which is at $\lambda=1$, $\beta=0$ and $M=\mu=0$. It is worth
noting that there are alternative relevant deformations we could have
considered involving the operators $\Tr\hat\Phi^{+2}$ and 
$\Tr\hat\Phi^{-2}$. These are equivalent to (\ref{tree}) 
as a deformation of the
${\cal N}=4$ theory, {\it i.e.}~when $\beta=0$, but differ once the
Leigh-Strassler marginal deformation is present. In particular, the
resulting theories differ in the IR. However, as they only differ by
strictly relevant operators they both flow to the same fixed point in
the UV: namely the Leigh-Strassler SCFT. The reason for choosing 
to study the specific relevant deformation (\ref{tree}) is simply that
the resulting matrix model turns out to be exactly soluble
\cite{DV,mm1,kostov}.

Aside from masses for the chiral multiplets $\hat\Phi^\pm$ and $\hat\Phi$,
our theory also involves two trilinear deformations,  
${\cal O}_1 =\Tr\hat\Phi[\hat\Phi^+,\hat\Phi^-]$ and ${\cal
O}_2 =\Tr\hat\Phi\{\hat\Phi^+,\hat\Phi^-\}$. 
Both the operators ${\cal O}_1$ and ${\cal O}_2$ are of course,
marginal by power counting but only one of them, or more precisely,
only one linear combination of these operators is an {\it exactly\/} 
marginal deformation of the $\N=4$ theory. 
Adding ${\cal O}_1$ to
the $\N=4$ Lagrangian only changes the coefficient in front of the $\N=4$
superpotential, but it is an irrelevant operator even at one-loop at the $\N=4$
fixed point
\cite{ofer}. The operator ${\cal O}_1$ is actually a descendant and
not a chiral primary field in the $\N=4$ theory, hence its
dimensions are not protected.\footnote{In the 
context of the AdS/CFT correspondence it is expected
that these operators will get mass dimensions $\sim (g^2_{YM}N)^{1/4}$
\cite{witten}
in the strongly coupled $\N=4$ theory in the large-$N$ limit.}
The second operator ${\cal
O}_2$ on the other hand is known to be exactly marginal at one-loop
at the $\N=4$ point \cite{ofer}. More generally, away from the $\N=4$
line one should expect, for fixed $g_{YM}$, a particular linear combination
of ${\cal O}_1$ and ${\cal O}_2$ to be exactly marginal. We
have parametrized this particular linear combination via two complex
numbers $\lambda$ and $\beta$. In principle, $\lambda$ should be
determined as a function of $\beta$ and $g_{YM}$ on the fixed
manifold. Fortunately the specifics of this will not be important for us. 
Note that in the $\SU(2)$
theory the operator ${\cal O}_2$ vanishes identically and the
above superpotential does not yield a marginal deformation. 
We will assume that $N>2$ throughout this paper.

The theory with masses for all the fields, as in Eq.~\eqref{tree}, has a
number of vacua, many of which are 
massive. The canonical
examples are the Higgs and confining vacua which will play an
important r\^ole in our discussion.
The confining vacua correspond to the trivial
classical solution of the $F$-term conditions
with $\hat\Phi=\hat\Phi^\pm=0$
and preserves the full $\SU(N)$ gauge symmetry classically. At low
energies the only classically massless fields comprise of the $\N=1$
gauge multiplet which confines and generates a mass gap. There is also
a Higgs vacuum where at the classical level the gauge symmetry is
completely broken. In addition, there are other massive vacua which are
visible classically as solutions that leave a non-abelian gauge
subgroup unbroken. Their classification is similar to that of the
$\N=1^*$ theory \cite{nick,VW,PS}. What is interesting about the
Leigh-Strassler deformation is that we shall find other vacua which
are not present in the $\N=1^*$ theory itself. In tandem with this,
for special values of the deformation parameter $\beta$, such that
$e^{i\beta}$ is a root of unity, vacua can disappear as a result of
the condensates diverging.

In this paper, we compute the values of the condensates
$\vev{\Tr\hat\Phi^k}$ and the
effective superpotential in all the massive vacua of the  $\SU(N)$
theory and we find that Montonen-Olive duality indeed relates the
holomorphic condensates in different vacua with a special action on
the couplings that we discuss below. This explicit computation is made
possible by the recent proposal of Dijkgraaf and Vafa \cite{DV} wherein the
effective superpotential of the gauge theory is computed by the genus
zero free energy of a holomorphic three-matrix integral
\cite{mm1}.
The matrix model however, computes the superpotentials
for the $\U(N)$ gauge theory. 
One of our important conclusions is that the
$\SU(N)$ superpotential differs non-trivially from its $\U(N)$
counterpart. (See also the work of \cite{ferrari} where a similar issue is  
discussed).
We show however that the former can be unambiguously
extracted from the latter by a process of ``integrating in'' of
additional fields present in the $\U(N)$ gauge theory.
This difference turns out to be crucial as it is the 
$\SU(N)$ results that clearly exhibit Montonen-Olive duality while the
$\U(N)$ results do not. For the $\SU(N)$ theory we find in a generic
$(p,k)$ massive vacuum (up to 
inconsequential vacuum-independent additive constants)
\EQ{W_{\rm eff}^{\SU(N)}={pN\mu M^2\over2\lambda^2\sin\beta}\cdot
{\theta_1^\prime(p\beta/2|\ttau)\over \theta_1({p\beta}/2
\vert{\ttau})}\ .
}
with
\EQ{\ttau={p\tau_R+k\over q};\quad\tau_R\equiv \tau-\frac{iN}\pi
\ln\lambda\ ;\qquad
p\cdot q=N\ ;\quad k=0,1,\ldots,q-1\ .}

The main consequence of this result is that the values of the
effective superpotential (and indeed all the eigenvalues of $\hat\Phi$) in 
different massive vacua of the theory with deformation parameter
$\beta$ are mapped into one another by the action of
the $\SL(2,{\mathbb Z})$ transformations on the couplings: 
\EQ{\tau_R\longrightarrow {a\tau_R+b\over c\tau_R+d}\ ;\qquad
\beta\longrightarrow{\beta\over{c\tau_R+d}}\ ; \qquad
\lambda^2\sin\beta\rightarrow {\lambda^2\sin\beta\over c\tau_R +d}\ ,}
$ad-bc=1$; $a,b,c, d\in {\mathbb Z}$. 
In particular, duality of the underlying SCFT is actually realized via
modular transformations on a particular combination of the
bare couplings $\tau_R$ rather than the gauge coupling $\tau$. In
addition, $\SL(2,\mathbb Z)$ permutes vacua of a theory with deformation
parameter $\beta$ provided $\beta$ transforms with modular weight $-1$
as above along 
with a specific action on $\lambda$ deduced from above. 

The vacuum structure and modular transformations described above can
also be understood via an exact elliptic superpotential for the theory
compactified on ${\mathbb R}^3\times S^1$. The superpotential is a
function of $N$ chiral superfields $X_a$ ($a=1,2, \ldots N$, 
$\sum_{a=1}^N X_a=0$) which
parameterize the Coulomb branch of the compactified theory. These
are a complex combination of the Wilson lines and
dual photons of the compactified theory. The
superpotential
\EQ{W_{{\mathbb R}^3\times S^1}={M^2\mu\over 2\lambda^2\sin\beta}\left({\omega_1\over\pi}
\right)\sum_{a\neq b}
\left(\zeta(X_a-X_b+\beta\omega_1/\pi)-\zeta(X_a-X_b-\beta\omega_1/\pi)
\right)
}
reproduces the vacua described in Eqs.~(1.3) and (1.4), taking the value
(1.3) described above. Here, $\zeta(z)$ is the Weierstrass-$\zeta$
function for the torus with half-periods $\omega_1=i\pi$ and
$\omega_2=i\pi\tau_R$, {\it i.e.\/}~with complex structure $\tau_R$. 
This superpotential also predicts 
new $\SL(2,{\mathbb Z})$-invariant vacua, not present in the $\N=1^*$
theory, whose existence is confirmed by the classical analysis of the
four-dimensional 
field theory. 

We also remark that the
properties of $\beta$ under modular transformations are
similar to those of individual gauge couplings in elliptic quiver
theories \cite{us,KV}. In addition, at non-zero $\beta$ the matrix
model solution naturally involves a torus with two marked
points with $\beta$ parameterizing the separation. 
These suggest a deeper connection between the
Leigh-Strassler and quiver SCFTs.

The layout of this paper is as follows: in Section 2 we show how one
can relate the superpotentials for the $\U(N)$ and $\SU(N)$ theories
by a process of integrating-in the trace part of the fields. This is
very important because the matrix model approach yields the
superpotential in the $\U(N)$ theory, whereas, we are primarily
interested in the $\SU(N)$ theory. In Section 3, following Dijkgraaf
and Vafa, we show how a matrix model can be used to find the
superpotential in the confining vacua of the $\U(N)$ theory and
therefore by implication in $\SU(N)$ theory. In particular, we find
the spectrum of eigenvalues of the adjoint-valued field $\hat\Phi$
from which all the holomorphic condensates can be calculated.
Section 4 is devoted to an analysis of the vacuum structure of the
mass deformed Leigh-Strassler theory from the point-of-view of the
tree-level superpotential. In particular, this leads to expressions
for the classical limit of the condensates in each of the massive vacua. In
addition, for the case of gauge group $\SU(3)$ we find new vacua that
are not present in the $\N=1^*$ theory. We also show how some of the
massive vacua can disappear when the coupling $\beta$ takes particular
values. In Section 5, we return to the matrix model and consider
multi-cut solutions that describe all the massive vacua. We then go on
to how the structure of vacua can
be used to deduce the action of the $\SL(2,{\mathbb Z})$ duality group
on the Leigh-Strassler theory itself. The final Section briefly
reports on how the results from the matrix model can be used to deduce
the exact elliptic superpotential (1.6) for the theory compactified on a
circle to three dimensions, generalizing the one for the $\N=1^*$
theory constructed in \cite{nick}. 

\section{Relation between the $\U(N)$ and $\SU(N)$ theories}

As we have explained one of our aims is to compute the quantum
superpotential and a certain set of holomorphic 
observables in the massive vacua of the mass deformed $\SU(N)$
Leigh-Strassler SCFT with $\N=1$ SUSY. For this purpose, we would like
to employ matrix model techniques that have been proposed recently by
Dijkgraaf and Vafa \cite{DV}.
However, already there is a subtlety. A direct
application of the matrix model approach will solve the $\U(N)$ theory
since the DV proposal relates the effective
superpotentials for $\U(N)$ $\N=1$ gauge theories to the planar diagram
expansion of corresponding matrix 
models. So the first question that we must address is
how are the $\SU(N)$ results related to those of the $\U(N)$ gauge
theory? We will now show that there is a very specific relation 
between the effective superpotential of the $\SU(N)$ theory with
classical superpotential Eq.~\eqref{tree}, and the effective
superpotential of the $\U(N)$ theory with the same classical
superpotential. This relation will eventually allow us to extract the $\SU(N)$
results from the matrix model for the $\U(N)$ gauge theory.

\subsection{The $\U(N)$ gauge theory}

The Leigh-Strassler theory with
$\SU(N)$ gauge group differs non-trivially from its $\U(N)$ counterpart. The
$\U(N)$ theory contains additional neutral chiral multiplets that couple to
the chiral superfields transforming in the adjoint representation of
the $\SU(N)\subset\U(N)$. These
interactions can modify the superpotential and other holomorphic
observables of the $\SU(N)\subset\U(N)$ theory. The way this happens is
clearly seen from the point-of-view of the $\U(N)$ theory.  Let
us begin by considering the $\U(N)$ version of the theory,
namely, the $\N=1$ SUSY gauge theory with tree
level superpotential
\EQ{W_{\rm cl}^{\U(N)}=\Tr\big(i\lambda\Phi[\Phi^+,\Phi^-]_\beta+M\Phi^+\Phi^-
+\mu\Phi^2 \big)\ ,\label{tree1}}
where $\Phi^\pm$ and $\Phi$ are the fields in the adjoint of
$\U(N)$. Now the fields $\Phi^\pm$ and $\Phi$ can be naturally split
into their traceless and trace parts:
\EQ{\Phi\equiv\hat\Phi+a\ ;\quad\Phi^\pm\equiv\hat\Phi^\pm+a^\pm\ 
;\qquad\Tr\hat\Phi\equiv
0\ ;\quad\Tr\hat\Phi^\pm\equiv0\ .}
Here, $\hat\Phi^\pm$ and $\hat\Phi$ are traceless and so transform in the adjoint
representation of $\SU(N)\subset\U(N)$ while $a^\pm=\Tr\Phi^\pm/N$ and
$a=\Tr\Phi/N$
are neutral. Rewritten in terms of these variables the tree level
superpotential \eqref{tree1} for the $\U(N)$ theory is
\EQ{W_{\rm cl}^{\U(N)}=N\big(M-2\lambda a \sin\tfrac\beta2\big)
a^+a^-+N\mu a^2+W_{\rm
cl}^{\SU(N)}(a,a^\pm)}
where
\SP{W_{\rm cl}^{\SU(N)}(a,a^\pm)=&
\Tr\left(i\lambda\hat\Phi[\hat\Phi^+,\hat\Phi^-]_\beta+\big(M-2\lambda a
\sin\tfrac\beta2\big)\,\hat\Phi^+\hat\Phi^-+\mu\hat\Phi^2\right.\\
&\left.\qquad\qquad-2 \lambda
\sin\tfrac\beta2\,a^-
\hat\Phi\hat\Phi^+-2\lambda\sin\tfrac\beta2\,a^+\hat\Phi\hat\Phi^-\right)\
.\label{tree2}
} 

The main point here is that the neutral trace fields $a$ and $a^\pm$ have
the effect of modifying
the couplings of the $\SU(N)$ fields. For example, the mass $M$ has been
renormalized to $M-2\lambda
a\sin\tfrac\beta2$. In addition, there are new bilinears in
$\hat\Phi^\pm$ and $\hat\Phi$ whose couplings\footnote{Strictly speaking, of
course, these ``couplings'' are actually chiral superfields. But from
the point of view the $\SU(N)$ sub-sector of the $\U(N)$ theory, these
neutral chiral superfields do appear like couplings that have been
elevated to chiral superfields otherwise known as ``spurions''.}
depend on $a,a^\pm$. (Notice that when $\beta=0$, that is for the $\N=1^*$ theory, 
these modifications disappear and there is no real difference between
$\SU(N)$ and $\U(N)$.)
In fact, we show below that because of the symmetries of the theory the VEVs
of $a^+$ and $a^-$ are forced to be zero self-consistently in the
full theory. Thus the only effect of the neutral fields is to modify
the mass term $M \rightarrow M^\prime\equiv
M-2a\lambda\sin{\tfrac\beta2}$. One might then suspect that answers for 
pure $\SU(N)$ theory of Eq.~\eqref{tree} may be obtained by simply
rescaling the $\U(N)$ results by appropriate powers of
$M^\prime=M-2a\lambda\sin{\tfrac\beta2}$. We will now see that this is
almost correct.  

First of all, the fields $a, a^\pm$ are blind to all gauge interactions
while $\hat\Phi,\hat\Phi^\pm$ experience
only $\SU(N)$ gauge interactions, the remaining $\U(1)$ gauge multiplet being
decoupled from everything else. Hence in a vacuum where the 
$\SU(N)$ fields are rendered massive and the gauge interactions
generate an effective superpotential for that sector of the theory, we
may readily write the effective superpotential for the $\U(N)$ gauge
theory as
\EQ{
W^{\U(N)}_{\rm eff}=N\big(M-2\lambda a \sin\tfrac\beta2\big)a^+a^-+N\mu
a^2+W_{\rm eff}^{\SU(N)}(a,a^\pm)\ .
\label{uneff}
}
Note that the chiral fields $a,a^\pm$ appear as parameters or
couplings for the $\SU(N)$ sub-sector, however they are actually
dynamical variables in the full theory and their values must be
determined by extremizing the full effective superpotential. 
This procedure can actually be implemented formally by first noting that
the $\SU(N)$ sub-sector, from Eq.~\eqref{tree2}, has certain abelian
symmetries. Let us define for the sake of convenience
$M^\prime\equiv M-2\lambda a \sin\tfrac\beta2$. It is then sufficient to
consider the following 
discrete symmetries: (i) $(\hat\Phi^\pm,a^\pm)$ $\rightarrow$
$-(\hat\Phi^\pm,a^\pm)$;
(ii) $(\hat\Phi^+,\hat\Phi,a^+, M^\prime,\mu)$ $\rightarrow$
$(-i\hat\Phi^+,i\hat\Phi,-ia^+,iM^\prime,-\mu)$; and (iii) $(\hat\Phi^-,\hat\Phi,a^-,
M^\prime,\mu)$ $\rightarrow$
$(-i\hat\Phi^-,i\hat\Phi,-ia^-,iM^\prime,-\mu)$.
The only possible form for the $\SU(N)$
effective superpotential, consistent with these symmetries, analytic
in the parameters and which has mass dimension three, is 
\EQ{W_{\rm eff}^{\SU(N)}=\mu M^{\prime2}\,F(a^+a^-/\mu M')\ ,\label{suneff}}
for some unknown function $F$. 
Plugging this back into Eq.~\eqref{uneff} for the effective
$\U(N)$ superpotential, and imposing the $F$-term conditions by
extremizing with respect $a,a^\pm$, the only solution that generates a
nontrivial effective superpotential is the one where 
\EQ{\langle a^\pm\rangle=0\,;\quad {\rm and} \quad  {\frac
{NM\mu}{2\lambda\sin\tfrac\beta2}}\,\langle a\rangle = W_{\rm
eff}^{\U(N)}\ .
\label{f}}
This tells us that in a vacuum of the $\U(N)$ gauge theory, the trace of the
adjoint scalar $\Phi$ must be related to the value of the effective
superpotential of the $\U(N)$ theory in that vacuum, precisely
according to the above equation. This already constitutes a non-trivial
prediction for the Dijkgraaf-Vafa approach for solving the $\U(N)$
gauge theory, one that our results from the matrix model must satisfy.

\subsection{The relation between the $\U(N)$ and $\SU(N)$ superpotentials}

The symmetry arguments above also imply that the effective
superpotential of the $\SU(N)$ theory with a classical action as
in Eq.~\eqref{tree} must be
\EQ{W^{\SU(N)}_{\rm eff}=\mu M^2\, F(0)\ .}
But we can easily determine the unknown function $F$ in terms of the
$\U(N)$ superpotential using Eqs.~\eqref{uneff}, \eqref{suneff} and
\eqref{f} and we find:
\EQ{W_{\rm eff}^{\SU(N)}=\frac{W_{\rm
eff}^{\U(N)}}{1-\frac{4\lambda^2\sin^2\beta/2}{N\mu M^2}
W_{\rm eff}^{\U(N)}}\ .
\label{rels}
}
Anticipating the matrix model results of the following section, we will
write this in a slightly different way that may turn out to be more
illuminating:
\EQ{
W_{\rm eff}^{\SU(N)}= \frac{M^2}{M^{\prime 2}(\vev{a})}\left[-W_{\rm
eff}^{\U(N)} +{N\mu M^2\over 4\lambda^2 \sin^2\tfrac\beta2}\right] -{N\mu
M^2\over 4\lambda^2 \sin^2\tfrac\beta2}\ .\label{relation}
}  
In this form, it is apparent that the effective superpotentials
of the $\U(N)$ and $\SU(N)$ gauge theories are indeed related by the
replacement $M^\prime(\vev{a})\rightarrow M$ after subtracting off certain
additive constants. This form of the relation will turn out to be
quite suggestive and useful when we discuss the matrix model results.
Importantly, this simple relation tells us how to extract the $\SU(N)$ answer
from
the $\U(N)$ result which the matrix model naturally computes. 

\section{The $\U(N)$ theory from the matrix model}

According to the Dijkgraaf-Vafa proposal \cite{DV}, in any given vacuum, the
effective superpotential of
the $\U(N)$ $\N=1$ gauge theory with classical superpotential
Eq.~\eqref{tree} is computed in terms of the planar diagram expansion of the
three-matrix model partition function expanded around that vacuum
\EQ{Z=\int\, [d\Phi^+]\,[d\Phi^-]\,[d\Phi]\,\exp-
g_s^{-1}\Tr\left(i\lambda\Phi[\Phi^+,\Phi^-]_\beta+M\Phi^
+\Phi^-+\mu\Phi^2\right)\ .
\label{mm}}
We use the same notation for the matrix fields and the associated
superfields in the $\U(N)$ theory. In the matrix model, unlike the
field theory, one takes $\Phi^+=(\Phi^-)^\dagger$ with the
fluctuations of $\Phi$ around the saddle point to be Hermitian. 
This matrix model has been actually solved in a different context
\cite{kostov} in the
large-$N$ limit. First one integrates out $\Phi^\pm$ and 
performs a field rescaling $\Phi\rightarrow\Phi/\lambda$ to get  
\EQ{Z=\lambda^{-N^2}\int\,[d\Phi]\,
{\frac{\exp -g_s^{-1}\mu\,\Tr\Phi^2/\lambda^2}
{\vert{\det(M{\bf 1}\otimes{\bf1}-i
e^{-i\beta/2}\Phi\otimes{\bf 1}+ie^{i\beta/2}{\bf
1}\otimes{\Phi})}\vert}}\ .\label{partfn}}
Now we can follow \cite{kostov} to obtain the saddle-point equation in
the large-$N$ limit. For completeness we will now follow the steps required.

Let $\{\phi_i\}$ denote the eigenvalues of the matrix $\Phi$. Changing
the integration variables in the matrix integral by going to the
eigenvalue basis introduces a Jacobian: the famous Van der Monde
determinant which leads to a repulsive force between the eigenvalues. 
The second step is a variable
change that will eventually yield a simplified form for the saddle-point
equation:
\EQ{\phi_i=-Me^{\delta_i}+{M\over2\sin\tfrac\beta2}\ .\label{change}}
In terms of these variables the {\it classical} potential for the
matrix model eigenvalues takes the following form
\EQ{\frac{\mu}{\lambda^2}\Tr\Phi^2=
\sum_i V(\delta_i)+{\mu
NM^2\over4\lambda^2\sin^2\tfrac\beta2}\ ,\label{classical}}
where we have defined
\EQ{V(\delta)\equiv{\mu
M^2\over\lambda^2}\left(e^{2\delta}-{e^{\delta}\over\sin\tfrac\beta2}\right)\
.} 
In addition to this classical potential, the eigenvalues $\delta_i$
also experience pairwise effective interactions induced both by the
Van der Monde determinant and the 
determinant resulting from integrating out $\Phi^\pm$ in
Eq.~\eqref{partfn}. The eigenvalues $\delta_i$ are naturally defined
on the
complex-$z$ plane with the identification $z\simeq z+2\pi i$, {\it
i.e.\/}~a cylinder.
 
\subsection{Solving the large-$N$ matrix model for the confining vacuum}

In the large-$N$ limit, the eigenvalues form a continuum and
condense onto 
cuts in the complex plane. On can think of these cuts as arising from
a quantum smearing-out of the classical eigenvalues of $\Phi$. For the
confining vacuum all the classical eigenvalues are degenerate, since
$\Phi\propto{\bf 1}$,  
and so we expect a solution in the matrix model involving 
a single cut. Multi-cut solutions will be discussed later. 
The extent of the cut and the matrix
model density of eigenvalues $\rho(\delta)$ can be determined from the
saddle-point
equation in terms of the parameters of the classical potential 
and the 't Hooft coupling of the matrix model $S=g_sN$. The
saddle-point equation is most conveniently formulated after defining the 
resolvent function:
\EQ{\omega(z)=\tfrac12\int_a^b d\delta\,
{\rho(\delta)\over\tanh{{z-\delta}\over 2}}\ ,\qquad
\delta\in[a,b]\ ,\qquad\int_a^b\rho(\delta)\,d\delta=1\ .\label{defres}} 
This function is analytic in $z$ and its only singularity is along a branch
cut extending between $[a,b]$. The matrix model spectral density
$\rho(\delta)$ is equal to the
discontinuity across the cut 
\EQ{\omega(\delta+i\epsilon)-\omega(\delta-i\epsilon)=-2\pi
i\rho(\delta)\ ;\qquad \delta\in[a,b]\ .}
The saddle-point equation expresses the zero force condition on a test
eigenvalue in the presence of the large-$N$ distribution of
eigenvalues along the cut. For the matrix model that we are studying,
this condition is best written in terms of the function $G(z)$ defined
as 
\EQ{G(z)=U(z)+i S\big(
\omega(z+i\tfrac\beta2)-\omega(z-i\tfrac\beta2)\big)\label{defg}}
and $\U(z)$ is a polynomial in $e^z$ such that 
\EQ{
V^\prime(z)=-i\big(U(z+i\tfrac\beta2)-U(z-i\tfrac\beta2)\big)\ .\label{defu}}
{}From this definition, one may easily deduce that $G(z)$ is an analytic
function on the cylinder and that in the interval $\vert {\rm
Im}z\vert\leq\pi$ it has two cuts 
$[a+i\tfrac\beta2,b+i\tfrac\beta2]$ and $[a-i\tfrac\beta2,b-i\tfrac\beta2]$. 
This is illustrated in Figure 1. 
\begin{figure}
\begin{center}\mbox{\epsfysize=5.5cm\epsfbox{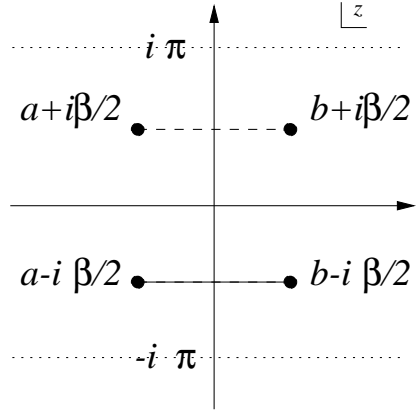}}\end{center}
\caption{\small The region over which the function $G(z)$ is defined
with its two cuts. The lines ${\rm Im}\,z=\pm i\pi$ are identified.}
\end{figure}
In terms of
$G(z)$, the matrix model saddle-point equation is 
\EQ{G(\delta+ i\tfrac\beta2\pm i\epsilon)=G(\delta-i\tfrac\beta2\mp
i\epsilon)\ ;\qquad\delta\in[a,b]\ .
\label{glue}}
The saddle-point equations actually provide a gluing condition on the
cylinder with two branch cuts. In particular, this defines a torus
with two marked points. $G(z)$ is then uniquely specified by gluing
condition Eq.~\eqref{glue} and asymptotic behaviour at
large $z$
\EQ{\lim_{z\rightarrow \infty}G(z)\rightarrow{\mu M^2\over
\lambda^2}\left[{1\over \sin\beta}e^{2z}-{1\over 2 \sin^2\tfrac\beta2}e^z
+ {\cal O}(e^{-z})\right]\ ,\label{asymp}}
which is a consequence of Eqs.~\eqref{defg} and
\eqref{defu}.
With this data we can determine $G(z)$ in a suitable parametrization
so that we can then implement the DV proposal and obtain the $\U(N)$
gauge theory superpotential.

\subsection{The $\ttau$-torus and the elliptic parametrization}

As we stated earlier, the two-cut cylinder along with the asymptotic
properties of $G(z)$ in Eq.~\eqref{asymp} uniquely specifies a torus
$E_\ttau$ with complex structure parameter $\ttau$. As shown in
Figure 2, the contour ${\cal C}_A$ enclosing one of the cuts
$[a+i\tfrac\beta2,b+i\tfrac\beta2]$
anticlockwise maps to the $A$-cycle of the torus while
the contour ${\cal C}_B$ joining the two cuts maps to the
$B$-cycle. We will think of this torus as a quotient of the complex
$u$-plane ($u$ being an auxiliary variable) by a lattice with periods
$2\tilde\omega_1$ and $2\tilde\omega_2$ and with complex structure
$\ttau=\tilde\omega_2/\tilde\omega_1$. 
\begin{figure}
\begin{center}\mbox{\epsfysize=5.5cm\epsfbox{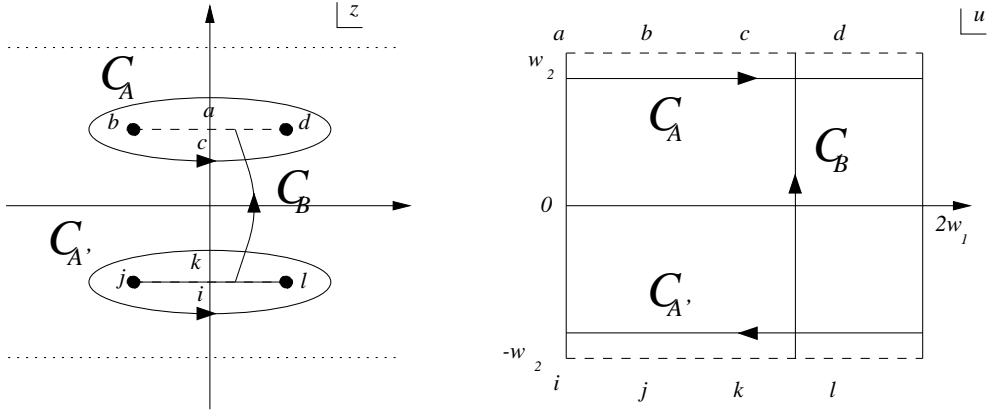}}\end{center}
\caption{\small Uniformizing map from the $z$-plane to the torus $E_\ttau$.}
\end{figure}

>From the field theory view-point, the appearance of this torus is not
entirely obvious except for the fact that the gauge theory we are
studying is connected to the $\N=4$ theory. In the $\N=1^*$ case
discussed in \cite{mm1,mm2}, one might have expected the torus to
appear due to the relation with the Seiberg-Witten curve of the
associated $\N=2^*$ theory. For the Leigh-Strassler SCFT the fact that
the torus structure is also present is one of the most important 
conclusions of this paper: as we shall see it leads to an
understanding of how $\SL(2,{\mathbb Z})$ acts on the underlying SCFT.
 
The map $z(u)$ from the $u$-plane to the $z$-plane is specified by the
requirements that going around ${\cal C}_A$ (the $A$-cycle of the
torus) returns
$z$ to its original value, while traversing the contour
${\cal C}_B$ (the $B$-cycle of the torus) causes $z$ to jump by an
amount $i\beta$, which is the distance between the two cuts in the
$z$-plane. Both these operations leave $G$ unchanged which is
therefore an elliptic function on the $u$-plane. Thus 
\AL{&A-{\rm cycle:}\quad z(u+2\tilde\omega_1)=z(u)\ ;\qquad
\qquad\ G(z(u+2\tilde\omega_1))=G(z(u))\ ,\\
&B-{\rm cycle:}\quad z(u+2\tilde\omega_2)=z(u)+i\beta\ ;\qquad
G(z(u+2\tilde\omega_2))=G(z(u))\ .
}
This determines the following unique map $z(u)$ from
the $u$-plane to the two-cut cylinder:
\EQ{\exp{z(u)}=\frac1{2\tan\tfrac\beta2}\cdot
{\theta_1^\prime(0\vert\ttau)\theta_1(\pi
u/2\tilde\omega_1-\beta/4\vert\ttau)\over\theta_1^\prime(\beta/
2\vert\ttau)\theta_1(\pi
u/2\tilde\omega_1+\beta/4\vert\ttau)}\label{defz}}
The fact that $G(z(u))$ is an elliptic function on the $u$-plane
along with its asymptotic behaviour at large $z$ \eqref{asymp} 
(corresponding to a
pole in the $u$-plane) leads uniquely to
\EQ{G(z(u))={\mu M^2\tilde\omega_1^2\cos\tfrac\beta2\over 4\pi^2
\lambda^2\sin^3\tfrac\beta2}
\cdot{\theta_1(\beta/2\vert\ttau)^2\over 
\theta_1^\prime(\beta/2\vert\ttau)^2}\,
\big[{\wp(u+\tilde\omega_1\beta/2\pi)-\wp(\tilde\omega_1\beta/\pi)}\big]\
.}
Here, $\wp(u)$ is the Weierstrass
function, an elliptic function defined on $E_\ttau$, 
and $\theta_1(u|\tau)$ is the Jacobi theta function of the first
kind (see \cite{ww} and Appendix A for more details). As expected,
$G(z(u))$ is an elliptic function of $u$ with a second order pole in
the $u$-plane; $z(u)$ on the other hand is only quasi-elliptic. For a
different choice of the deformation potential the asymptotic behaviour of
$G(z)$ will change and $G(z(u))$ will end up being an elliptic
function of a different order, but significantly 
$z(u)$ will remain unchanged. There
are two special points on the torus $E_\ttau$ at
$u=\pm\beta\tilde\omega_1/2\pi$ which map to the points $z=\mp
\infty$. Having determined $G(z)$ in the elliptic parametrization we
can now implement the DV proposal to compute the superpotential in the
confining vacuum of the $\U(N)$ gauge theory.

\subsection{Superpotential of the $\U(N)$ theory via Dijkgraaf-Vafa}

According the Dijkgraaf-Vafa proposal, the gluino condensate of the gauge
theory gets identified with the 't Hooft coupling $S$ of the matrix
model. From Eqs.~\eqref{defres} and \eqref{defg}, the integral of  $G(z)$ around
one
of its cuts, say ${\cal C}_A$, is equal to $-2\pi S$, so we can directly compute
the
gluino condensate of the $\U(N)$ gauge theory in the elliptic
parametrization:
\EQ{\Pi_A=2\pi i S = -i\int _A G(z(u)){dz(u)\over du}du\ .}
The second ingredient required to determine the QFT
superpotential is the variation in the genus zero free energy of the
matrix model in transporting a test eigenvalue from infinity to the
endpoint of the cut. This is obtained by integrating the force on a
test eigenvalue from infinity to the endpoint of the cut and can
be expressed as $\Pi_B=-i\int_B G(z) dz$. There is a
slight subtlety here as there is a ``zero-point'' energy associated
with additive constants in the matrix model potential. First note that
there is a multiplicative factor of $\lambda^{-N^2}$ in the
partition function \eqref{partfn} and an additive constant in the
classical scalar potential \eqref{classical}. Both these appear as 
$S$-dependent contributions to the matrix model genus zero free energy
defined as ${\cal F}_0/g_s^2=N^2{\cal F}_0/S^2$, so that 
\EQ{{\partial {\cal F}_0\over \partial S}=\Pi_B-2S\ln\lambda+{\mu
M^2\over 4\lambda^2\sin^2\tfrac\beta2}\ ;\qquad
\Pi_B= -i\int _B G(z(u)){dz(u)\over du}du\ .
\label{dfds}} 
The effective
superpotential of the mass deformation of the $\U(N)$ Leigh-Strassler
theory in its confining vacuum is obtained by extremizing the following
expression with respect to $S$:
\EQ{W_{\rm eff}^{\U(N)}=N{\partial {\cal F}_0\over \partial S}-2\pi i
\tau S;\qquad {\partial \over \partial S}{W_{\rm eff}^{\U(N)}}=0\ .}
where $\tau$ is the bare coupling of the theory. The $S$-independent
constant in Eq.~\eqref{dfds} appears as a vacuum-independent additive
constant in the superpotential. On the other hand, the term linear in $S$ in
Eq.~\eqref{dfds} actually 
has the effect of renormalizing the bare coupling $\tau$ so that 
\EQ{\tau\ \longrightarrow\ \tau_R = \tau-{iN\over\pi}\ln\lambda\ .
\label{tren}
}

Both $\Pi_A$ and  $\Pi_B$ are easy to evaluate in our elliptic
parametrization:
\EQ{\Pi_A= {d h(\ttau)\over d\ttau}\ ;\qquad \Pi_B=\ttau {d
h(\ttau)\over d\ttau}-h(\ttau)\ ,}
where
\EQ{
h(\ttau)={\mu M^2\cos\tfrac\beta2\over 4\lambda^2\sin^3\tfrac\beta2}\cdot
{\theta_1(\beta/2|\ttau)\over\theta_1^\prime(\beta/2\vert\ttau)}\ 
.\label{defh}}
It follows straightforwardly that 
\EQ{\frac{\partial}{\partial S} W_{eff}^{\U(N)}=0\quad  \Longrightarrow
\quad \ttau={\tau_R\over N}}
so that
\EQ{W_{\rm eff}^{\U(N)}=-Nh(\tau_R/N)+{N\mu
M^2\over4\lambda^2\sin^2\tfrac\beta2}\ .}
Finally we have, therefore, the quantum superpotential of the $\U(N)$
gauge theory in its confining vacuum,
\EQ{W_{\rm eff}^{\U(N)}=-{N\mu M^2\cos\tfrac\beta2\over
4\lambda^2\sin^3\tfrac\beta2}\cdot\frac{\theta_1(\beta/2|\tau_R/N)}
{\theta'_1(\beta/2|\tau_R/N)}+{N\mu M^2\over
4\lambda^2\sin^2\tfrac\beta2}\ ,\label{un}}  
with $\tau_R$ defined in \eqref{tren}. This result satisfies the first
obvious check: in the $\beta\rightarrow 0$ limit with
$\lambda=1$ it reproduces the superpotential of the $\N=1^*$
theory. The superpotential in the other $N-1$ confining vacua are
obtained by repeated $T$-dualities $\tau_R\to\tau_R+1$. Note that the
$T$-dualities in question naturally act on $\tau_R$ rather than the
bare coupling $\tau$. Similarly the result for the superpotential has
a modular parameter $\tau_R/N$ and so we expect that any
modular properties we uncover should naturally be a consequence of
the modular group acting on $\tau_R$ the ``renormalized'' coupling.

\subsection{Eigenvalues of $\Phi$ in the $\U(N)$ theory and a test}

One of the main results of our earlier work \cite{mm1} in the context
of $\N=1^*$ theory was that the map $z(u)$ from the
$u$-plane to the $z$-plane, when evaluated along the $A$-cycle of the
$\ttau$-torus (which maps to the upper cut on the $z$-plane), precisely
yields the eigenvalues of the adjoint scalar $\Phi$ of the gauge
theory in its confining vacuum, distributed uniformly along this 
cycle.\footnote{In contrast the matrix model eigenvalues are not distributed
uniformly along this cycle signalling the crucial difference between
the spectral densities of the two systems. Nevertheless, the matrix model
provides a means to extract the field theory distributions in a simple way.}
We were able to prove this assertion
for a class of deformations of the $\N=4$ theory involving
masses for two of the adjoint chiral superfields and arbitrary polynomials in
the third adjoint field $\Phi$. We will not pursue this generalized
analysis in the present case, but we will present strong evidence that
the same relation with the field theory eigenvalues holds under the
Leigh-Strassler deformation.

For the Leigh-Strassler theory, we first recall that the $\delta_i$
are valued in the $z$-plane 
and these are in turn related to
the eigenvalues of the $\Phi$ matrix through
Eq.~\eqref{change}. Taking into 
account the fact that the function $G(z)$ is defined 
in Eq.~\eqref{defg} in terms of the resolvent functions
$\omega(z\pm i\tfrac\beta2)$ where the positions of the cuts of the latter
have been displaced by $\pm i\tfrac\beta2$, we conjecture that
the function $-M\exp{(z(u)-i\tfrac\beta2)}/\lambda$ evaluated along the
$A$-cycle gives the eigenvalues of $\Phi$ in the $\U(N)$ theory:
\EQ{\phi(x)={M\over2\lambda\sin\tfrac\beta2}
\left[-\cos\tfrac\beta2\,{\theta_1^\prime(0\vert\tau_R/N)
\over\theta_1^\prime(\beta/2\vert{\tau_R/N})}
\;\;{\theta_3(\pi x -\beta/4\vert{\tau_R/N})
\over\theta_3(\pi x +\beta/4\vert{\tau_R/N})}\;+1\right]\,;\qquad
x\in[-\tfrac12,\tfrac12]\ .\label{LSeigen}
}
Note that the factor $\lambda$ takes into account the field rescaling
prior to Eq.~\eqref{change} and we also incorporate the additive shift
of $M/(2\lambda\sin\tfrac\beta2)$ 
involved in the variable change in Eq.~\eqref{change}. 
This is of course a continuous distribution of eigenvalues appropriate
for the $\U(N)$ theory in the large-$N$ limit. But the finite $N$
answer is obtained by simply replacing the continuous index $x$ by a
discrete index. For this to be legitimate, it is essential that the
field theory eigenvalues be uniformly distributed in the 
interval $x\in[-\tfrac12, \tfrac12]$ as we have indeed pointed out earlier.
The difference between finite-$N$ and large-$N$ manifests
itself in gauge-invariant observables as vacuum-independent operator mixing
ambiguities that vanish in the large-$N$ limit \cite{mm1}. 

Our results for the eigenvalues in the confining vacuum of the $\U(N)$
theory satisfy several checks. Firstly in
the classical limit $\tau_R\rightarrow i\infty$, using the properties
of the Jacobi theta functions (see Appendix A), Eq.~\eqref{LSeigen} vanishes as
expected, since the confining vacuum should correspond to the trivial
classical solution. Secondly, in the $\beta\rightarrow 0$ limit with 
$\lambda=1$ the eigenvalues of 
$\N=1^*$ theory in its confining vacuum reproduce the expressions in 
\cite{mm1,DS}.
Two further nontrivial tests involve the computation of $\vev{\Tr\Phi}$ and
$\vev{\Tr\Phi^2}$ in the $\U(N)$ theory using the eigenvalues
above. With a large-$N$ distribution, the trace reduces to an integral
and we find
\AL{&\Vev{\Tr\Phi}=N\int_{-1/2}^{1/2} dx\,\phi(x)=
{2\lambda\sin\tfrac\beta2\over \mu
M}\,W_{\rm eff}^{\U(N)}\ ,\label{tr1}\\
&\Vev{\Tr\Phi^2}=N\int_{-1/2}^{1/2} dx\,\phi^2(x)={1\over
\mu}\,W_{\rm eff}^{\U(N)}\ ,\label{tr2}}
where $W_{\rm eff}^{\U(N)}$ is the expression obtained in
Eq.~\eqref{un}. We show how to compute the associated integrals
explicitly in Appendix B.
The first \eqref{tr1} demonstrates that the trace of $\Phi$ calculated
using the matrix model results  satisfies precisely the relation
Eq.~\eqref{f} predicted on
general grounds from field theory. The second integral is also exactly
in accord with field theory expectations since from our general
arguments in Section 2, we have  $W_{\rm eff}^{\U(N)}\propto \mu$ and
therefore differentiating with respect to $\mu$ gives 
$\vev{\Tr\Phi^2}=W^{\U(N)}_{\rm eff}/\mu$ precisely \eqref{tr2}.

In summary, the eigenvalues of the adjoint scalar $\Phi$ in the
confining vacuum of the mass deformed $\U(N)$ Leigh-Strassler theory
at large $N$ are
given by Eq.~\eqref{LSeigen}. The eigenvalues of the $\U(N)$
theory at finite $N$ are obtained by substituting the continuous index $x$
with an appropriate discrete index.

\subsection{The $\SU(N)$ superpotential}

We argued in Section 2 that the superpotentials of the $\U(N)$ and 
$\SU(N)$ theories differ non-trivially, but are nevertheless simply
related by Eq.~\eqref{relation}. The main point there was that
the $\U(N)$ theory allows a non-zero $\vev{\Tr\Phi}\equiv N\vev{a}\neq 0$ to
develop in the confining vacuum of the quantum theory. This leads to
what is effectively a renormalization of the mass parameter
$M\rightarrow M^\prime=M-2\vev{a}\lambda\sin\tfrac\beta2$. According to 
the result of our field theory analysis of Section 2, 
the $\SU(N)$ superpotential may be recovered by a 
rescaling of the $\U(N)$ result by $M^2/M^{\prime2}$, after subtracting
off a vacuum independent constant. It is quite interesting to note  
that precisely the same vacuum independent constant makes an appearance
in the $\U(N)$ superpotential \eqref{un} computed in the
Dijkgraaf-Vafa approach. 

Using our results for the $\U(N)$ superpotential from the previous
section and Eq.~\eqref{relation}, we
thus find that in the  confining vacuum the $\SU(N)$ theory develops the
superpotential,
\EQ{W_{\rm eff}^{\SU(N)}={N\mu M^2\over2\lambda^2\sin\beta}
\cdot{\theta_1^\prime(\beta/2|\tau_R/N)\over
\theta_1(\beta/2\vert\tau_R/N)}\,-{N\mu M^2\over4\lambda^2\sin\tfrac\beta2}\ ,}
where the renormalized coupling is defined in \eqref{tren}.
This expression for the $\SU(N)$ superpotential has the correct
$\beta\rightarrow 0$ limit in that it reproduces the $\N=1^*$
superpotential exactly.

\subsection{$\SU(N)$ eigenvalues and condensates}

The eigenvalues $\hat\phi_i$ of the adjoint scalar $\hat\Phi$ in the confining
vacuum of the $\SU(N)$ theory are related to the $\U(N)$ eigenvalues by a simple
extension of our arguments in
Section 2 as  
\EQ{\hat\phi(x)={M\over
M^\prime}\Big(\phi(x)
-{M\over2\lambda\sin\tfrac\beta2}\Big)+{M\over2\lambda\sin\tfrac\beta2}\
.}
This can be understood simply as a rescaling of the $\U(N)$ spectrum
of eigenvalues by $M/M^\prime(\vev{a})$ after subtracting out a
constant and is in line with the arguments of Section 2 for extracting
the $\SU(N)$ superpotential.

Evaluating this explicitly yields
\EQ{\hat\phi(x)
={M\over2\lambda}
\left[-{\theta_1^\prime(0\vert\tau_R/N)
\over\theta_1(\beta/2\vert\tau_R/N)}
\;\;{\theta_3(\pi x -\beta/4\vert\tau_R/N)
\over\theta_3(\pi x +\beta/4\vert\tau_R/N)}\;+\;{1\over
\sin\tfrac\beta2}\right];\qquad
x\in[-\tfrac12,\tfrac12]\ .\label{confining}}
We show in Appendix B that with these eigenvalues $\vev{\Tr\hat\Phi}=0$
and $\vev{\Tr\hat\Phi^2}=W_{\rm eff}^{\SU(N)}/\mu$. Once again we remark
that with a continuous label $x$ these are really the eigenvalues of
the large-$N$ gauge theory but finite-$N$ results are obtained by
simply replacing $x$ with a discrete index.

Now that we have the complete distribution of eigenvalues of $\hat\Phi$ in
the confining vacuum of the $\SU(N)$ theory, every condensate of the form
$\langle\Tr\hat\Phi^k\rangle$
is automatically computed via the integrals 
\EQ{\Vev{\Tr\hat\Phi^k}=N\int_{-1/2}^{1/2}dx\, \hat\phi^k(x)\ .}
We will not undertake the exercise of computing these but, if desired, they
can be computed following the methods in Appendix B.

\subsection{Summary}
In this section we have computed the value of the
effective superpotential and the spectrum of eigenvalues of $\hat\Phi$
in the confining vacuum of the $\SU(N)$
theory. We did this by obtaining the $\U(N)$ superpotential via a
matrix model and subsequently extracting the $\SU(N)$ results by a process of
integrating-in\footnote{We remark that one might have expected to
obtain the $\SU(N)$ superpotential and eigenvalues directly from the
matrix model by imposing a constraint on the matrices via a Lagrange
multiplier. However this turns out not to work in any simple way, the
primary reason being, as pointed out in \cite{mm1}, that the matrix
model eigenvalue spectrum is completely different from the field
theory spectrum and tracelessness of the latter translates into a
rather complicated constraint on the former.} discussed in Section
2. The results of the computation have 
satisfied several general field theory checks. In addition, the emergence of a
quasi-modular function ($\theta_1^\prime/\theta_1$) as
the superpotential of the $\SU(N)$ theory is 
strongly suggestive of Montonen-Olive duality being realized in this
theory. (In contrast, the $\U(N)$ superpotential being
basically the reciprocal of the $\SU(N)$ result, transforms in a
complicated way under the modular group.) In order to establish
electric-magnetic duality we need to be able to compute the
corresponding quantities in other massive vacua of the
theory (most notably the Higgs vacuum which should be $S$-dual to the
confining phase). We will address this in Sections 4 and 5 which will
be devoted to understanding the 
classical and quantum vacuum structure respectively.

\section{Vacuum structure of deformed $\SU(N)$ Leigh-Strassler theory}

In this section we exhibit the classical vacuum structure of the
$\SU(N)$ theory. Importantly these classical configurations will serve
as the starting point for analysing other saddle points of the matrix
model and thence obtaining the effective quantum superpotential in all
the massive vacua. Interestingly the classical solutions (the
eigenvalues) for the Higgs vacuum that we find here will be shown to
be related by $S$-duality to the quantum eigenvalue spectrum of the
confining vacuum that we have already obtained in the previous section.

The vacuum structure of the mass deformed $\SU(N)$
Leigh-Strassler theory turns out to be subtly different from that of
the $\N=1^*$ theory: there are new vacua and old vacua can disappear. 
In order to describe the vacuum structure, we start with the 
tree-level superpotential \eqref{tree}. At the classical level vacua
are determined by solving the 
$F$- and $D$-flatness conditions modulo gauge
transformations. As usual, if one is interested in holomorphic
information it is sufficient to solve the
$F$-flatness conditions modulo {\it complex\/} gauge
transformations. These conditions follow from extremizing the
tree-level superpotential \eqref{tree}:\footnote{Note that the
$F$-term equations are rather similar to the 
``$q$-deformed'' $\SU(2)$ algebra. They are similar to, but not the same as the
equations encountered in \cite{LB}. The authors of \cite{LB}
considered a slightly different mass deformation and a $\U(N)$ gauge group.}
\EQ{
{\mathbb P}\,[\hat\Phi^+,\hat\Phi^-]_\beta=-i\frac{2\mu}\lambda\hat\Phi\ ,\qquad
{\mathbb
P}\,[\hat\Phi,\hat\Phi^+]=-i\frac 
M\lambda\hat\Phi^+\ ,\qquad
{\mathbb
P}\,[\hat\Phi^-,\hat\Phi]=-i\frac M\lambda\hat\Phi^-\ .
\label{fflatness}
}
In the above, ${\mathbb P}$ is the projector onto the traceless part which
must be imposed since we are dealing with the $\SU(N)$, rather than
the $\U(N)$, theory. Complex gauge transformation can be used to 
diagonalize $\hat\Phi$.

In the limit $\beta=0$, the theory becomes the basic $\N=1^*$ theory
and the conditions \eqref{fflatness} reduce to the Lie algebra of
$\SU(2)$. Let us briefly recall the solutions in that case. Each
solution is associated to a representation of $\SU(2)$ of dimension
$N$ where $\hat\Phi$ and $\hat\Phi^\pm$ are, up to a trivial re-scaling, $J_3$
and $J_\pm$, respectively. Each representation of $\SU(2)$ of
dimension $N$ is associated to a partition $N\to n_1+\cdots+n_q$,
where each integer $n_i$ is the dimension of each irreducible
component. In general the unbroken gauge group has abelian factors and
the phase is Coulomb, or massless. The massive vacua are associated to
the the equi-partitions $N=q\cdot p$ for which the unbroken gauge
group is $\SU(q)$. In such a vacuum, the low energy description is
pure $\N=1$ with gauge group $\SU(q)$ which has $q$ inequivalent
quantum vacua. Hence, there are $\sum_{q|N}q$ vacua with a mass gap (the 
sum being over the positive integer divisors of $N$).

Now we consider the case when $\beta\neq0$. Each of the vacua of the
${\cal N}=1^*$ theory remains a vacuum of the deformed theory, at least for
generic values of $\beta$. These deformed solutions have the following
form. Firstly, the block form of the solutions associated to the
partition $N\to n_1+\cdots+n_q$ is preserved. Let us concentrate on 
one of the blocks of dimension $n$. 
For $\beta=0$, $\hat\Phi$ and $\hat\Phi^\pm$ are equal to
$J_3$ and $J_\pm$, up to trivial re-scaling, of the irreducible
$\SU(2)$ representation of dimension $n$:
\EQ{
\beta=0\ :\qquad\hat\Phi=-\frac M\lambda J_3\
;\quad\Phi^\pm=\frac{2\mu M}\lambda J_\pm\ .
} 
When $\beta\neq0$, the non-zero elements of 
$\hat\Phi^\pm$ have the same pattern as $J_\pm$ (so only the elements just above and below the
diagonal, respectively are non-zero) but the numerical values are changed. The
elements of diagonal matrix $\hat\Phi$ in the block in question are found
to be 
\EQ{
\hat\phi_{j}=\frac{M}{2\lambda}\Big(\frac1{\sin\tfrac\beta2}-\frac
n{\sin\tfrac{n\beta}2}
e^{-i\beta(n+1-2j)/2}\Big)\ ,\qquad j=1,\ldots,n\ .
\label{eigb}
}
Notice that when $\beta\to0$, $\hat\Phi$, in this block, becomes equal to 
$-(M/\lambda)J_3$ as expected. 

For generic $\beta$, these deformed vacua are massive/massless
according to their type when $\beta=0$. However, for specific values
of $\beta$ vacua can migrate off to infinity in field space 
and hence cease to be genuine vacua of the 
field theory. This
occurs when the eigenvalues of $\hat\Phi$ diverge. It is easy to see that
this will occur in a block of dimension $n$ when the deformation parameter 
\EQ{
\beta=\frac{{2\pi k}}n\ ,\qquad k\in{\mathbb Z}\ .
\label{spb}
}

In one of the $N/p$ massive vacua with unbroken $\SU(N/p)$ symmetry
the eigenvalues of $\hat\Phi$ are given by \eqref{eigb} with $n=p$ and
where each eigenvalue has degeneracy $N/p$. From this spectrum, we can calculate
the classical values for the condensates; for instance, for the
quadratic condensate:
\EQ{
\langle{\rm Tr}\,\hat\Phi^2\rangle=\frac{M^2N}
{4\lambda^2\sin^2\tfrac\beta2}\Big(p\frac{
\tan\tfrac\beta2}{\tan\tfrac{p\beta}2}-1\Big)\ .
\label{csl}
}
Notice that this condensate diverges when $\beta=2\pi k/p$,
$k\in{\mathbb Z}$ in accord with \eqref{spb}.

\subsection{New classical branches}
 
As well as the original massive vacua,
there are also new vacua which are not present when
$\beta=0$. We illustrate this is in the simplest case where the gauge
group is $\SU(3)$. In this case there is a new vacuum, more precisely
a moduli space of vacua, in which $\hat\Phi$ and $\hat\Phi^\pm$ are
all, up to complex gauge transformations, diagonal:
\SP{
\hat\Phi&=\frac M{2\lambda\sin\tfrac\beta2}\,{\rm diag}\big(x_1,x_2,x_3\big)\ ,\\
\hat\Phi^+&=\frac{\sqrt{\mu M/2}}{\lambda\sin\tfrac\beta2}\,\rho\,{\rm diag}
\big((1-x_1)^{-1},(1-x_2)^{-1},(1-x_3)^{-1}\big)\ ,\\
\hat\Phi^-&=\frac{\sqrt{\mu M/2}}{\lambda\sin\tfrac\beta2}\,\xi\,{\rm diag}
\big((1-x_1)^{-1},(1-x_2)^{-1},(1-x_3)^{-1}\big)\ .
}
Here, $\rho$ and $\xi$ are two complex moduli of the solution and $x_i$,
$i=1,2,3$ are the three distinct roots of the cubic
\EQ{
x^3-3x+2-\rho\xi=0\ .
\label{cub}
}
Note that the tracelessness of $\hat\Phi$ and $\hat\Phi^\pm$ is guaranteed 
by the form of the cubic \eqref{cub}. 
Clearly these solutions diverge to infinity when $\beta\to0$. However,
the moduli space of the solutions also has additional branches which
occur whenever 
\EQ{
e^{i\beta/2}x_i-e^{-i\beta/2}x_j=2i\sin\tfrac\beta2\ .
}
In this case, an off-diagonal component of either of $(\hat\Phi^\pm)_{ij}$ can
become non-zero. The quadratic condensate in this space of vacua is
\EQ{
\SU(3):\qquad
\langle{\rm Tr}\,\hat\Phi^2\rangle=\frac{3M^2}{2\lambda^2\sin^2\tfrac\beta2}\ .
\label{new}
}
Clearly this diverges when $\beta=0$. We will see that these new
vacua are important for checking a conjecture that we make later. 

\section{Electric-magnetic duality in the Leigh-Strassler Deformation}

The question we now address is what
becomes of the $S$-duality of the $\N=4$ theory under the Leigh-Strassler
deformation? Specifically we are considering a deformation of the
$\N=4$ Lagrangian of the form; 
\begin{equation}
{\cal L}={\cal L}_{{\cal N}=4} + \int\,d^{2}\theta\, \Delta W 
\end{equation}
where the deforming superpotential can be written as 
\begin{equation}
\Delta W= i\big(\lambda\cos\tfrac{\beta}{2}-1\big) {\cal O}_{1}
-\lambda\sin\tfrac{\beta}{2}{\cal O}_{2} +
M\,\Tr\hat\Phi^{+}\hat\Phi^{-}+ \mu\,\Tr\hat\Phi^{2} 
\end{equation}
where, as in Section 1, 
${\cal O}_1 =\Tr\hat\Phi[\hat\Phi^+,\hat\Phi^-]$ and ${\cal 
O}_2 =\Tr\hat\Phi\{\hat\Phi^+,\hat\Phi^-\}$. At the ${\cal N}=4$
point, the operators appearing in the deforming superpotential
transform under $S$-duality in a definite way. Adding these operators 
to the superpotential therefore breaks modular invariance. However, as
is standard when dealing with broken symmetries, we should be able to 
restore invariance by allowing the corresponding couplings to
transform in an appropriate way. This strategy was implemented in
detail for the ${\cal N}=1^{*}$ case
$\beta=0$ in \cite{ADK}. We will now review this briefly and extend
this discussion to the marginal deformation.  

Using the conventions 
described in \cite{ADK},\footnote{See the final paragraph of Section 3
of this reference.} the superspace measure $d^{2}\theta$ has
holomorphic modular weight $-2$. Hence for modular invariance of the 
Lagrangian, the
superpotential must have weight $+2$.   
In the $\N=4$ theory, the mass operators 
$\Tr\hat{\Phi}^{+}\hat{\Phi}^{-}$ and $\Tr\hat{\Phi}^{2}$ are 
both part of the multiplet of chiral
primary operators which transform as second rank symmetric
traceless tensors under the $\SO(6)$ R-symmetry. Their modular weights
can be deduced via the AdS/CFT correspondence by comparison with the
dual fields in IIB supergravity \cite{int}. In fact 
they have holomorphic modular weight $+2$ and the Lagrangian will be
invariant if we assign the mass parameters $M$ and $\mu$ weight zero. 
Although there is no explicit breaking of modular invariance, it is
important to remember that the theory has various vacuum states in
which these mass operators acquire expectation values, thereby
`spontaneously' breaking $S$-duality. In other words, modular
transformations map the observables in one vacuum to those in another.
Indeed, this property is manifest in the exact results of
\cite{nickprem,ADK}.  

Similar arguments should apply to the Leigh-Strassler
theory at least near the $\N=4$ point where the modular properties of
the deforming operators are known. If we set $\lambda=1$ and work to 
linear order in $\beta$, 
the deformation only contains the cubic operator ${\cal O}_{2}$. This
operator is also part of an $\N=4$ chiral
primary multiplet consisting, in this case, of third rank symmetric
traceless tensors of $SO(6)$. In the conventions of \cite{ADK} it
has modular weight $+3$. To restore invariance under $S$-duality we
should therefore assign the parameter $\beta$ holomorphic
modular weight $-1$. In Section 5.2, we will find the appropriate
generalization of this transformation property throughout the
parameter space, by using the matrix model to find
the exact superpotential in each of the massive vacua. In Section 5.1
we see what can be deduced by looking at the Higgs and confining vacua alone.

\subsection{A first look at $S$-duality}
Let us concentrate on the Higgs and confining vacua.
In the $\N=1^*$ limit, it is well known that each observable in
the Higgs vacuum is simply 
related by $S$-duality, $\tau\rightarrow-1/\tau$, 
to the corresponding observable
in the confining vacuum. This is an explicit 
realization of the Montonen-Olive electric-magnetic duality of the
$\N=4$ theory. The question is whether a similar duality persists under the
Leigh-Strassler deformation.

If $S$-duality is
realized in this theory it should exchange the Higgs and confining vacua.
Since we already have the eigenvalues in the confining vacuum of the
$\SU(N)$ theory a simple and powerful test emerges: use $S$-duality to
deduce the eigenvalues in the Higgs vacuum and then take the classical
limit and compare with the spectrum \eqref{eigb} ($n=N$) deduced from the
$F$-flatness conditions. In the next section, we shall use the matrix
model to solve for the superpotential in each of the massive vacua 
and this uncovers the action of the whole of the modular group.

The eigenvalues of $\hat\Phi$ in the confining vacuum
are given in \eqref{confining}. (Recall that the eigenvalues in this
vacuum have the property that they vanish in the classical limit.)
The first thing to notice is that the modular parameter of the elliptic
function parametrizing the eigenvalue spectrum is $\tau_R/N$ with
$\tau_R=\tau-(iN/\pi)\ln\lambda$, where $\tau$ is the gauge
coupling. Hence $S$-duality, if
realized, must act on the renormalized coupling $\tau_R$ as
$\tau_R\rightarrow-1/\tau_R$ rather than $\tau$ itself. 
Under this action, the modular parameter
of the elliptic functions changes as
$\tau_R/N\rightarrow-1/(\tau_RN)$. But now we want to take the 
classical limit $\tau_R\rightarrow i\infty$ and this can be
done easily after performing a modular transformation $-1/(\tau_R
N)\rightarrow \tau_R N$ on the (quasi-)elliptic functions. 
Using the modular transformation properties of
the theta functions (see Appendix A) we find 
\EQ{
\hat\phi(x)\ \longrightarrow\ -\tau_R{NM\over
2\lambda\sin\tfrac{N\tau_R\beta}2}
e^{-iN\tau_R\beta x}+{M\over2\lambda\sin\tfrac\beta2};\quad
x\in[-\tfrac12,\tfrac12]\ .
} 
Apart form the additive constant,
this coincides precisely with the expression Eq.~\eqref{eigb} ($n=N$) for
the eigenvalues in the Higgs vacuum of the deformed 
$\SU(N)$ Leigh-Strassler theory with $\beta$ replaced by $\tau_R\beta$. 
Equivalently, the Higgs and confining vacua of the
theory with deformation parameter $\beta$ are simply exchanged under the
combined operation $\tau_R\rightarrow-1/\tau_R$ and $\beta\rightarrow\beta/\tau_R$

The overall factor of $\tau_R$ in the above
expression is expected because the eigenvalues have modular weight
one. The fact that the additive constants are different is to be
expected: even in the context of $\N=1^*$ theory the matrix model approach
only yields results that are $S$-duality
covariant up to a vacuum independent shift \cite{mm1}. This additive
shift will be discussed in more detail in the next section. We should
also point out that after $S$-duality on the spectrum of eigenvalues
in the confining vacuum 
we have taken a classical limit above. In this limit, the exact
transformation properties of the couplings $\lambda, \beta$ are not
visible. These will become clear in the next section.

The results of this simple analysis are that there is indeed a form of 
$S$-duality that relates the Higgs and confining vacua of the
theory with deformation parameter $\beta$. In particular this duality  
acts on the bare couplings of the Leigh-Strassler theory as 
\EQ{
\tau_R\ \longrightarrow\ -1/\tau_R\ ,\qquad
\beta\ \longrightarrow\ \beta/\tau_R\ .
\label{sdu}
}  
This is consistent with $\beta$ having modular weight -1 as expected.
Recall that the bare theory has 3 couplings $\tau$, $\lambda$ and
$\beta$ with one non-trivial relation, so that \eqref{sdu} determines
the action of $S$-duality on the bare theory up to the unknown
relation. In the next section we expand the discussion of duality to
include the whole of the modular group $\SL(2,{\mathbb Z})$ and deduce
the action on all three of the bare couplings.

\subsection{Multi-cut solutions of the matrix model and $\SL(2,{\mathbb Z})$}

Now that we have described the structure of massive vacua of the
Leigh-Strassler deformed theory, we can return to the matrix model
and ask whether we can solve for the exact superpotential and
condensates in each of these vacua. This will turn out to be very important
because it will prove that there is an action of electric-magnetic
duality on the Leigh-Strassler deformation of the $\N=4$ theory. In
fact this extends to the action of the full $\SL(2,{\mathbb Z})$ modular
group. 

The way to use the matrix model to solve for the theory in an
arbitrary massive vacuum is described for the $\N=1^*$ theory in
\cite{mm2}. One finds that the same techniques can be used in the
deformed matrix model described in Section 3. In order to describe the
massive vacua with unbroken $\SU(N/p)$ symmetry, one needs a multi-cut
solution of the matrix model. Of course the matrix model computes the
superpotential in the $\U(N)$ theory and we then use \eqref{rels} to
deduce it in the $\SU(N)$ theory.

We start with the observation that the
classical eigenvalues in the massive vacua with unbroken $\SU(N/p)$
are given by \eqref{eigb} with $n=p$, each with degeneracy $N/p$. The
only thing that will be important for us will be the range of the
phases of these classical eigenvalues. The classical eigenvalues of
the $\U(N)$ gauge theory also have exactly the same phases in the corresponding
massive vacua.  As
we have seen explicitly in the case of the confining vacuum, the eigenvalues of
the field theory are essentially proportional to $\exp[z]$. 
This
directs us to consider a $p$-cut solution of the matrix model with the
cuts in the $z$-plane situated at 
\EQ{
\big[a+i\beta(p+1-2j)/2,b+i\beta(p+1-2j)/2]\ ;\qquad j=1,\ldots,p\,
}
where these positions are dictated precisely by the phases of the
classical eigenvalues. The classical eigenvalues are clustered in
groups of $N/p$ at each of these points, but the pairwise effective
interactions induced in the matrix model will cause each cluster to spread
out into a cut.

Following \cite{mm2}, we then make an ansatz where the cuts have the same
length and the same filling fractions. It can then be shown that
$G(z)$ defined in \eqref{defg} again  only 
has two cuts, as in the confining vacuum, but where the cuts are
shifted to $\pm[a+i\beta p/2,b+i\beta p/2]$. This modifies the map
$z(u)$ in \eqref{defz} by a shift $\beta\to p\beta$:
\EQ{\exp{z(u)}=\frac1{2\tan\tfrac{p\beta}2}\cdot
{\theta_1^\prime(0\vert\ttau)\theta_1(\pi
u/2\tilde\omega_1-p\beta/4\vert\ttau)\over\theta_1^\prime(p\beta/
2\vert\ttau)\theta_1(\pi
u/2\tilde\omega_1+p\beta/4\vert\ttau)}\ .\label{defz2}}
Similarly
\EQ{G(z(u))={\mu M^2\tilde\omega_1^2\cos\tfrac{p\beta}2\over 4\pi^2
\lambda^2\sin^3\tfrac{p\beta}2}
\cdot{\theta_1(p\beta/2\vert\ttau)^2\over 
\theta_1^\prime(p\beta/2\vert\ttau)^2}\,
\big[{\wp(u+\tilde\omega_1p\beta/2\pi)-\wp(\tilde\omega_1p\beta/\pi)}\big]\
.}

When the Dijkgraaf-Vafa recipe is followed one finds that for each cut
there is an associated filling fraction of eigenvalues $N_j/N$ 
which translates to an
effective 't Hooft coupling $S_j=g_s N_j$ with $\sum S_j=S$. For the
massive vacua we note that the filling fractions in each of the $p$
cuts are the same and equal to $N/p=q$
\EQ{
2\pi iS_j={d h(\ttau)\over d\ttau}\ ;\qquad j=1,\ldots,p\ ,
}
where $h(\ttau)$ is the same function that we encountered in \eqref{defh} with $\beta\to p\beta$:
\EQ{
h(\ttau)={\mu M^2\cos\tfrac{p\beta}2\over 4\lambda^2\sin^3\tfrac{p\beta}2}\cdot
{\theta_1(p\beta/2|\ttau)\over\theta_1^\prime(p\beta/2\vert\ttau)}\ .
}
In addition one also needs the variation in the genus zero free energy
upon transporting an eigenvalue from any given cut to infinity. After minor modifications this
is also given by the same expressions that were encountered in the
one-cut solution for the confining vacuum: 
\EQ{\sum_{j=1}^p{\partial {\cal F}_0\over \partial S_j}=\ttau {d
h(\ttau)\over d\ttau}-h(\ttau)-2\ln\lambda\sum_{j=1}^pS_j+p{\mu
M^2\over 4\lambda^2\sin^2\tfrac{\beta}2}\ .
\label{dfdsn}
}
Then the superpotential for the $\U(N)$ theory at any of its massive
vacua is
\EQ{W_{\rm eff}^{\U(N)}=\frac Np\sum_{j=1}^p
{\partial {\cal F}_0\over \partial S_j}-2\pi i
\tau\sum_{j=1}^p S_j\ .
}
Extremizing with respect to $\ttau$ gives
\EQ{
\ttau=\frac{p^2}N\tau_R
}
and therefore we find that the value of the quantum effective
superpotential for the $\U(N)$ theory is 
\EQ{
W_{\rm eff}^{\U(N)}=-{N\mu M^2\cos\tfrac{p\beta}2\over
4p\lambda^2\sin^3\tfrac{p\beta}2}\cdot\frac{\theta_1(p\beta/2|p^2\tau_R/N)}
{\theta'_1(p\beta/2|p^2\tau_R/N)}+{N\mu M^2\over
4\lambda^2\sin^2\tfrac{\beta}2}\ .\label{unn}
}  
In all of the above we should remember that $p$ is a divisor of $N$.

The result for $\SU(N)$ follows by the application of
Eq.~\eqref{rels}:
\EQ{
W_{\rm eff}^{\SU(N)}={pN\mu M^2\over2\lambda^2\sin\beta}\cdot
{\theta_1^\prime(p\beta/2|p^2\tau_R/N)\over \theta_1({p\beta}/2
\vert{p^2\tau_R/N})}\;-{N\mu M^2\over4\lambda^2\sin^2\frac\beta2}\ .
\label{exv}
}
This gives the result in only one of $q=N/p$ vacua of this type. The
results in the remaining vacua follow by the replacement 
\EQ{
\tau_R\longrightarrow\tau_R+k/p\ ;\qquad k=0,1,\ldots q-1\ .
}
The Higgs vacuum corresponds to $p=N$ whilst the confining vacuum
considered previously has $p=1$ and $k=0$. 

One interesting and immediate check on the
form of the superpotential follows from the fact that it is manifestly
divergent when the deformation parameter takes the special values
\EQ{
\beta=\frac{2\pi k}p\ ,\qquad k\in{\mathbb Z}\ .
}
But this matches precisely the values of $\beta$ in
Eq.~\eqref{spb} for which the classical eigenvalues of $\hat\Phi$ for
the massive vacua diverge (since they consist of $q$ blocks of size $n=p$).
Another check follows by taking the classical limit of the  
superpotential \eqref{exv}. Using \eqref{scl}, one finds
\EQ{
W_{\rm eff}^{\SU(N)}\ \overset{\tau_R\to i\infty}\longrightarrow\ 
\langle{\rm Tr}\,\hat\Phi^2\rangle=\frac{N\mu M^2}
{4\lambda^2\sin^2\tfrac\beta2}\Big(p\frac{
\tan\tfrac\beta2}{\tan\tfrac{p\beta}2}-1\Big)\ ,
}
which matches the values of $\mu\langle{\rm Tr}\,\hat\Phi^2\rangle$
obtained in \eqref{csl} from the $F$-flatness condition precisely.

Now that we have the superpotential in each of the massive vacua,
we can now investigate how the full $\SL(2,{\mathbb Z})$ duality group
acts on the theory. In the $\N=1^*$ theory described in \cite{mm1,mm2},
we established that the matrix model yields a result for the
superpotential which is only duality covariant when shifted by a
vacuum independent constant. This kind of additive ambiguity is
ubiquitous in
this subject: see \cite{mm1,nick}.
Naturally, the same kind of ambiguity will be present in the
Leigh-Strassler deformation where we claim that the appropriate
definition of a duality covariant superpotential is
\EQ{
\tilde W_{\rm eff}^{\SU(N)}=W_{\rm eff}^{\SU(N)}+C(\tau_R,\beta,\lambda)
\ ,
}
where the vacuum-independent constant is given by either of the expressions
\SP{
C(\tau_R,\beta,\lambda)
&=\frac{N\mu M^2}{4\lambda^2\sin^2\frac\beta2}-
\frac{N\mu
M^2}{2\lambda^2\sin\beta}\cdot\frac{\theta'_1(\beta/2|\tau_R)}
{\theta_1(\beta/2|\tau_R)}
+\frac{N(N-1)\mu M^2\beta}{12\lambda^2\sin\beta}E_2(\tau_R)\\
&=\frac{N\mu M^2}{4\lambda^2\sin^2\frac\beta2}-
\frac{N\mu
M^2\omega_1}{\pi\lambda^2\sin\beta}\Big(\zeta(\omega_1\beta/\pi)-
\frac{N\beta}\pi
\zeta(\omega_1)\Big)\ .
\label{defc}
}
Here, the Weierstrass $\zeta$-function is defined relative to a torus
with periods $\omega_1$ and $\omega_2$ and $\tau_R=\omega_2/\omega_1$
and $E_2(\tau_R)$ is the $2^{\rm nd}$ Eisenstein series (see the
Appendix of \cite{mm1} for definitions and references). One property that the
$C$ has is that it reduces to the correct form needed in the
$\N=1^*$ theory in the limit $\beta\to0$.

Our claim is that the duality-covariant definition of the 
superpotential in the each of the massive vacua
labelled by $(p,k)$, $p\cdot q=N$ and $k=0,\ldots,q$, is then
\EQ{
\tilde W_{\rm eff}^{\SU(N)}={N\mu M^2\over2\lambda^2\sin\beta}\Big(
p{\theta_1^\prime(p\beta/2|\ttau)\over \theta_1({p\beta}/2
\vert{\ttau})}-{\theta_1^\prime(\beta/2|\tau_R)\over \theta_1({\beta}/2
\vert{\tau_R})}+\frac\beta6(N-1)E_2(\tau_R)\Big)\ ,
\label{hhhp}
}
where
\EQ{
\ttau=(p\tau_R+k)/q\ .
}

Using the modular properties of the theta functions and the $2^{\rm
nd}$ Eisenstein series, \eqref{modu1} and \eqref{modu2}, one
can easily show, for example, that
\EQ{
\tilde W_{\rm eff}(\beta/\tau_R,-1/\tau_R,\lambda)\Big\vert_{\rm Conf}=
\tau_R{\sin\beta\over\sin(\beta/\tau_R)}\cdot
\tilde W_{\rm eff}(\beta,\tau_R,\lambda)\Big\vert_{\rm
Higgs}\ .
}
Notice that the coupling $\beta$ transforms with unit weight 
under the modular group. Note also the rather unconventional
transformation of 
the superpotential itself: this we will have to cure by an appropriate
definition of the transformation of the coupling $\lambda$. 

A consideration of the other massive vacua show that they all lie on a
single orbit of the $\SL(2,{\mathbb Z})$ duality group. For the
element
\EQ{
\sigma=\MAT{a&b\\ c&d}\in\SL(2,{\mathbb Z})\ ,
} 
which relates two vacua $A=\sigma(B)$, we have\footnote{The actual
action of $\sigma$ on a vacuum with labels $(p,k)$ is not difficult to
elucidate.}
\EQ{\tilde W_{\rm eff}\big(\tfrac\beta{c\tau_R+d},\tfrac{a\tau_R+b}{c\tau_R+d},
\lambda\big)\Big\vert_{A}=
(c\tau_R+d)\frac{\sin\beta}{\sin[\beta/(c\tau_R+d)]}
\cdot\tilde W_{\rm eff}(\beta,\tau_R,\lambda)\Big\vert_{B}\ .
}
{}From this we can deduce the action of $\sigma\in\SL(2,{\mathbb Z})$ on
the bare couplings of the theory; clearly
\EQ{
\sigma(\tau_R)=\frac{a\tau_R+b}{c\tau_R+d}\ ,\qquad
\sigma(\beta)=\frac\beta{c\tau_R+d}\ ,
\label{trans}
}
confirming that $\beta$ transforms with modular weight -1. However,
we can say more: in order that the superpotential transforms with a
definite modular weight, so that it is consistent with the group
action, we can deduce the transformation of the coupling
$\lambda$. Recall, that on the marginal surface of the
Leigh-Strassler deformation
$\lambda$ is a function---albeit unknown---of
the other two couplings and so must transform. In addition, since in
the limit $\beta\to0$, the superpotential transforms with modular
weight 2, we expect this to be preserved for $\beta\neq0$. This
requires the combination $\lambda^2\sin\beta$ that appears as an
overall factor in the superpotential must transform with unit modular
weight:
\EQ{
\sigma(\lambda^2\sin\beta)=\frac{\lambda^2\sin\beta}{c\tau_R+d}\ .
}

\section{An exact elliptic superpotential}

Another method that has been used to calculate the exact values of the
superpotential in all the vacua of the $\N=1^*$ theory is to
compactify it on a circle of finite radius \cite{nick}. The effective
superpotential is then a function of the dual photons and Wilson lines
of the abelian subgroup $\U(1)^{N-1}\subset\SU(N)$. These comprise
$N-1$ complex scalar fields $X_a$, $a=\ldots,N$ (with
$\sum_{a=1}^NX_a=0$) which naturally live on a torus of complex
structure $\tau$ because of the periodicity of each dual photon and
Wilson line. The superpotential 
describing the $\N=1^*$ deformation is therefore constrained to be an elliptic
function of the complex scalars $X_{a}$. In \cite{nick} the exact
answer was found to be   
\EQ{
W_{{\mathbb R}^3\times S^1}\thicksim\sum_{a\neq b}\wp(X_a-X_b)\ ,
\label{ees}
}
where $\wp(z)$ is the Weierstrass function defined on a torus with
half periods $\omega_{1,2}$ with $\tau=\omega_2/\omega_1$. What is
particularly useful about this superpotential is that it is
independent of the compactification radius, and therefore, yields
results that are valid in the four-dimensional limit, and it encodes
all the vacua of the $\N=1^*$ theory: both massive and massless.

The question is whether there is an elliptic superpotential that
describes the Leigh-Strassler deformation? Similar arguments to those
used in \cite{nick} suggest that there should be, but are not
powerful enough to determine the exact answer. However, there is an
obvious candidate for appropriate marginal deformation of (\ref{ees}),
which turns out to be correct. Specifically, we can replace the 
Weierstrass function $\wp(z)$, by its unique deformation in the
space of even elliptic functions.\footnote{The function
must be even to respect the Weyl group of $\SU(N)$} In particular, the
Weierstrass function has one double pole at the origin and we can
consider the elliptic function of order two obtained by splitting the
double pole into two single poles. Remarkably, a superpotential of
precisely this form reproduces our results from the matrix model 
for the superpotential in each of the massive vacua. This
superpotential can be written as,  
\EQ{W_{{\mathbb R}^3\times S^1}={M^2\mu\over 2\lambda^2\sin\beta}\left({\omega_1\over\pi}
\right)\sum_{a\neq b}
\left(\zeta(X_a-X_b+\beta\omega_1/\pi)-
\zeta(X_a-X_b-\beta\omega_1/\pi)
\right)\ .
\label{esp}
}
The Weierstrass-zeta function $\zeta(u)$ can be defined via
$\wp(u)=-\zeta^\prime(u)$ where $\wp(u)$ is the Weierstrass function
for the torus with complex structure $\tau_R$. 
Then in the limit $\beta=0$ this reproduces \eqref{ees}. What is
remarkable is that the massive vacua
of the $\N=1^*$ theory for which the $X_a$ lie on sub-lattices are also
critical points of \eqref{esp} since this only relies on the fact that the
two-body interaction is an even elliptic function.  For instance, for
the $p\cdot q=N$ massive vacuum with $k=0$ we have
\EQ{
X_a\in\Big\{\frac{2\omega_1r}p+\frac{2\omega_2s}q\ ,\quad
r=0,\ldots,p-1\ ,\quad s=0,\ldots,q-1\Big\}\ .
}
Furthermore, the values of the
superpotential are precisely given by \eqref{hhhp}. 

A very strong check on the form of the superpotential we have
conjectured in \eqref{esp} is possible
because of the existence of new vacua which are not present in the
limit $\beta=0$. For example, it is easy to show that $X_a=0$, $\forall
a$, is a critical point. Actually this is an $\SL(2,{\mathbb
Z})$-invariant vacuum. Now we can take the classical limit,
$\tau\to i\infty$, after having shifted by minus the vacuum-independent
shift \eqref{defc}, and compare with the analysis of the vacuum
structure in Section 4. In the case of $\SU(3)$,  the 
classical limit of the superpotential $W_{{\mathbb
R}^3\times S^1}-C$ in the vacuum with
$X_a=0$ precisely equals the value calculated from the new vacuum
solution of the $F$-flatness conditions \eqref{new}. We expect this to
generalize to $\SU(N)$, however, the analysis remains to be done.

The properties and construction of this elliptic superpotential will
be described in a separate publication; however, we mention
the fact that for $\N=1^*$, the elliptic superpotential is, more or
less, the (complexified) 
Hamiltonian of the elliptic Calogero-Moser integrable system
which described $N$ particles interacting though pair-wise forces
proportion to $\wp(X_a-X_b)$. The intriguing question is whether this
system can be deformed
\EQ{
\wp(X_a-X_b)\to\zeta(X_a-X_b+\beta\omega_1/\pi)-
\zeta(X_a-X_b-\beta\omega_1/\pi)
}
whilst preserving integrability.

Finally we comment on a remarkable correspondence between our results
for the Leigh-Strassler deformation and the exact superpotential of a
certain relevant deformation of the $A_{1}$-quiver theory
studied in \cite{us}. Specifically this is the theory of an  
$\SU(N)\times\SU(N)$ gauge multiplet with gauge couplings $g_{1}$ and
$g_{2}$, with two chiral multiplets $A$ and
$B$ in the $(N,\bar{N})$ representation and two chiral multiplets
$\tilde{A}$ and $\tilde{B}$ in the $(\bar{N},N)$. The theory also contains
two chiral multiplets $\hat\Phi_{1}$ and $\hat\Phi_{2}$ in the $({\rm adj},0)$
and $(0,{\rm adj})$ representations respectively with classical 
superpotential
\begin{equation}
W= {\rm Tr}\left\{(g_{1}\hat\Phi_{1}+M/2)(A\tilde{A}+B\tilde{B})+
(g_{2}\hat\Phi_{2}+M/2)(\tilde{A}A+\tilde{B}B)\right\}\ .
\end{equation}
This theory actually has ${\cal N}=2$ supersymmetry but we further 
deform it by adding the operator, 
\EQ{
\mu{\rm Tr}\,\big(\hat\Phi_1^2-\hat\Phi_2^2\big)\ .
}
to the classical superpotential which preserves ${\cal N}=1$
supersymmetry. This theory is a relevant deformation of the ${\cal
N}=2$ superconformal theory obtained by setting $\mu=M=0$. 
The exact vacuum structure and superpotential was
determined in \cite{us}. The theory has $N\sum_{d|N}d$ massive vacua. 
Remarkably, in $\sum_{d|N}d$ of 
these vacua, we find exactly the same value of the
superpotential as in each of the massive vacua of the $\SU(N)$
Leigh-Strassler theory studied in this paper. Specifically, this
agreement holds if we identify  
the parameter $\beta$ of the
Leigh-Strassler deformation and the complexified coupling of one of the
$\SU(N)$ factors in the quiver theory $\tau_1=4\pi
i/g_{1}^{2}+\theta_{1}/2\pi$ (or $z=2i\pi\tau_1$ to use the notation of
\cite{us}) :
\EQ{
z=\frac{N\omega_1\beta}{\pi}\ .}
In the brane set-up for the quiver model $z$ is the relative
separation of the two NS5-branes. We also identify the parameters $M$
and $\mu$ of the quiver superpotential with parameters of the same
name in the deformed Leigh-Strassler superpotential (\ref{tree}).   

This connection suggests that there should be an RG flow from the 
superconformal $A_{1}$-quiver to the Leigh-Strassler fixed line. 
At least in one special case, such a flow is known to exist. 
In \cite{KW}, Klebanov and Witten considered the quiver theory in the case of
equal gauge couplings $g_{1}=g_{2}$ and zero masses for the
bi-fundamental multiplets, $M=0$. In this case one can integrate out
the adjoint chiral fields $\hat\Phi_{1}$ and $\hat\Phi_{2}$ and induce a 
marginal quartic superpotential for the bi-fundamentals which
corresponds to D3-branes placed at a resolved conifold
singularity. Moving onto the Higgs branch where $\SU(N)\times\SU(N)$ is
broken to the diagonal $\SU(N)$, we find an effective theory of three 
chiral superfields transforming in the adjoint of the unbroken $\SU(N)$
which coincides with ${\cal N}=4$ SUSY Yang-Mills. Repeating this
exercise with unequal gauge couplings $g_{1}\neq g_{2}$, we find that
the Leigh-Strassler marginal deformation as well as other irrelevant operators
are induced. However, it is not obvious whether this flow is related to the
exact agreement we have found above. 

\vspace{1cm}

{\bf Acknowledgements}: TJH would like to thank Frank Ferrari for
discussions. The authors would also like to thank Robbert Dijkgraaf
and Cumrun Vafa for valuable discussions. We thank Robbert Dijkgraaf
for his comments and observations on a preliminary version of this article.
SPK acknowledges support from a PPARC Advanced Fellowship.

\startappendix

\Appendix{Some properties of elliptic functions}

The (quasi-)elliptic functions that we need, $\wp(z)$, $\zeta(z)$ and
$\theta_i(z|\tau)$, are all standard and their definitions and properties
can be found in \cite{ww}. In our conventions, the basic torus has
periods $2\omega_1$ and $2\omega_2$ with complex structure
$\tau=\omega_2/\omega_1$. We also define $q=\exp(i\pi\tau)$. Below we
list some properties of these functions that will be useful.

An important relation between the Weierstrass and Jacobi theta functions is
\EQ{
\zeta(u)-\frac{\zeta(\omega_1)}{\omega_1}u=\frac{\pi}{2\omega_1}
\frac{\theta'_1(\pi u/(2\omega_1)|\tau)}
{\theta_1(\pi u/(2\omega_1)|\tau)}\ .
\label{relt}
}
The Jacobi theta functions have the following modular properties that will be
useful
\AL{
\theta_1(x|\tau)&=i(-i\tau)^{-\tfrac 1
2}\exp(-ix^2/\pi\tau)\theta_1(x/\tau|-1/\tau)\ ,\label{modu1}\\
\theta_3(x|\tau)&=(-i\tau)^{-\tfrac 1
2}\exp(-ix^2/\pi\tau)\theta_3(x/\tau|-1/\tau)\ .
\label{modu3}
}
In order to take the classical limit, we will need the expansion
\AL{
\theta_1(x|\tau)&=2\sum_{n=0}^\infty(-1)^nq^{(n+1/2)^2}\sin(2n+1)x\
,\\
\theta_3(x|\tau)&=1+2\sum_{n=0}^\infty(-1)^nq^{n^2}\cos2nx\
,\\
\frac{\theta'_1(x|\tau)}{\theta_1(x|\tau)}&={\rm cot}\,x+4\sum_{n=1}^\infty
\frac{
q^{2n}}{1-q^{2n}}\sin2nx\ .\label{scl} 
}

We also need the $2^{\rm nd}$ Eisenstein series \cite{Kob}
\EQ{
E_2(\tau)=1-24\sum_{n=1}\sigma_1(n)q^{2n}\ ,
}
where $\sigma_1(n)$ is a sum over each positive integral divisor of
$n$. It has the following non-trivial modular transformation
\EQ{
E_2(-1/\tau)=\tau^2E_2(\tau)+\frac{6\tau}{\pi i}\ .
\label{modu2}
}

\Appendix{Calculating condensates}

In this appendix we show how the integrals of the type \eqref{tr1} and
\eqref{tr2} can be computed to yield the condensates in the $\U(N)$ and
$\SU(N)$ gauge theories. Let us first consider the function $f(x)$ in
terms of which the eigenvalues have been defined
\EQ{f(x)={\theta_4(\pi x-\beta/4\vert\ttau)\over\theta_4(\pi
x+\beta/4\vert\ttau)}\equiv{\theta_3(\pi x-\pi/
2-\beta/4\vert\ttau)\over\theta_3(\pi
x-\pi/2+\beta/4\vert\ttau)};\quad x\in[0,1]\ .}
Clearly, the eigenvalues can be written either in terms of $\theta_3$
or $\theta_4$ since the only difference is in the range of the
argument $x$. We also define the function $g(x)$
\EQ{g(x)={\theta_1(\pi x-\beta/4\vert\ttau)\over\theta_1(\pi
x+\beta/4\vert\ttau)}\ .}
The functions $f$ and $g$ have some nice properties
\EQ{g(x+1)=g(x)\ ;\quad g(x+\ttau/2)=e^{-i\beta/2}f(x)\ ;\quad
g(x-\ttau/2)=e^{i\beta/2}f(x)\ ,}
which follow from standard identities.
Note also that the function $g(x)$ has a simple pole in its
fundamental period at $x=1-\beta/4\pi$. Now it is a simple matter to
convince oneself that because of the periodicity properties of
$g(x)$ the contour integral around this pole and hence the residue
there can also be written as
\EQ{{\rm
Res}\;g(x)\Big\vert_{x=-\beta/4\pi}=-2i\sin\tfrac\beta2\,\int_0^1f(x)\,dx\
.}
Thus we see that 
\EQ{\int_0^1 f(x)\,dx={1\over
\sin\tfrac\beta2}{\theta_1(\beta/2|\ttau)\over\theta_1^\prime(0|\ttau)}\
.\label{f1}}
{}From this key result it follows using Eq.~\eqref{LSeigen}
\EQ{\Vev{\Tr\Phi}=N\int_{0}^{1}dx\,\phi(x)={NM\over2\lambda\sin\tfrac\beta2}
\left[-{1\over\tan\tfrac\beta2}{\theta_1(\beta/2|\ttau)
\over\theta_1^\prime(\beta/2|\ttau)}
+1\right]
={2\lambda\sin\tfrac\beta2\over M\mu}W_{\rm eff}^{\U(N)}\ .}
(Note that because $\phi(x)$ is a periodic function under $x\rightarrow
x+1$ the actual limits of integration can always be changed from
$(-1/2,1/2)$ to $(0,1)$.)
Similarly it also follows that in the $\SU(N)$ theory
\EQ{\Vev{\Tr\hat\Phi}=N\int_{0}^{1}\hat\phi(x)\,dx=0\ .}

To compute the quadratic condensates we need the integrals
$\int_0^1f^2(x)dx$. These can also be calculated using the same trick
as above and we find that
\EQ{\int_0^1f^2(x)\,dx=-{1\over 2i\sin\beta}{\rm
Res}\;g^2(x)\Big\vert_{x=-\beta/4\pi}=
\frac{2\theta_1^\prime(\beta/2|\ttau)
\theta_1(\beta/2|\ttau)}{\sin\beta\,\theta_1^{\prime}(0|\ttau)^2}\ .
\label{f2}}

Using the results Eqs.~\eqref{f1} and \eqref{f2} we can readily
compute the quadratic condensates. For the $\U(N)$ theory we find
\EQ{\Vev{\Tr\Phi^2}=N\int_{0}^{1}\phi^2(x)\,dx
=-{NM^2\over 2\lambda^2\sin\beta}\cdot{\theta_1(\beta/2|\ttau)\over
\theta_1^\prime(\beta/2|\ttau)}+{NM^2\over 4\lambda^2\sin^2\tfrac\beta2}={1\over\mu}W_{\rm eff}^{\U(N)}\ .} 
For the $\SU(N)$ theory we find similarly
\EQ{\Vev{\Tr\hat\Phi^2}=N\int_{0}^{1}\hat\phi^2(x)\,dx
={ NM^2\over2\lambda^2\sin\beta}\cdot
{\theta_1^\prime(\beta/2|\ttau)\over
\theta_1(\beta/2\vert\ttau)}-{NM^2\over4\lambda^2\sin^2\tfrac\beta2}=
{1\over\mu}W_{\rm eff}^{\SU(N)}}
Similarly higher condensates could be computed explicitly if required
and will in general depend on the residue of $g^{n}(x)$ at the pole
$x=-\beta/4\pi$

\end{document}